\documentclass[12pt]{iopart}
\usepackage{graphicx}
\begin{document}

\title[Heavy Cosmic Ray Nuclei from Extragalactic Sources above ''The Ankle'']
{Heavy Cosmic Ray Nuclei from Extragalactic Sources above ''The Ankle''}

\author{Tadeusz Wibig$^1$ and Arnold W. Wolfendale$^2$}

\address{$^1$Physics Department, University of Lodz, 
Lodz, Poland\\$^2$ Physics Department, Durham University, Durham, UK}
\ead{wibig@zpk.u.lodz.pl}
\begin{abstract}
A very recent observation by the Auger Observatory group \cite{[1]} claims strong evidence for cosmic rays above 5.6$\times 10^{19}$ eV - (56 EeV) being protons from Active Galactic Nuclei. If, as would be expected, the particles above the ankle at about 2 EeV are almost all of extragalactic (EG) origin then it follows that the characteristics of the nuclear interactions of such particles would need to be very different from conventional expectation -- a result that follows from the measured positions of 'shower maximum' in the Auger' work. 

Our own analysis gives a different result, viz that the detected particles are still 'massive' specifically with a mean value of $\langle \ln A \rangle = 2.2
\pm 0.8$. The need for a dramatic change in the nuclear physics disappears. 
\end{abstract}

\pacs{96.50.sb, 96.50.sd, 98.70.Sa}
\maketitle

\section{Introduction}

\begin{figure}[bh]
\centerline{
\includegraphics[width=15cm]{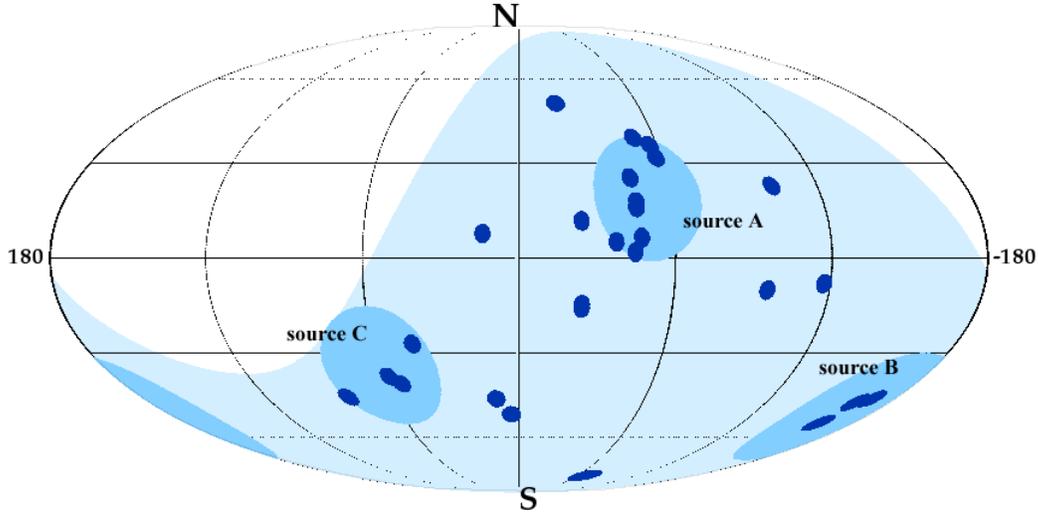}}
\caption{AUGER source map showing possible 'sources' A, B and C
\cite{[1]}. 
The energy threshold is 57 EeV.\label{fig1}}
\end{figure}

It has long been suggested that the particles above the ankle are extragalactic (e.g. \cite{[2],[3]}); indeed, some believe that the transition starts at an even lower energy than 2 EeV (e.g. \cite{[4]}). There have been many claims for EG 'signals' from specific sources (e.g. \cite{[5]}) but, apart for rather strong evidence for particles from the VIRGO cluster (the centre of the supercluster in which we are situated) the results have been conflicting. There were thus high expectations for the results from the very large Auger Observatory and such result, based on an exposure (area times time) exceeding the sum total of the world's data have recently appeared \cite{[1]}. Figure~\ref{fig1} shows the results and it is evident that there is 'clumpiness' in the arrival directions. The authors draw attention to a number of features, principally

\begin{itemize}
\item[1]The presence of a cluster of events round the CEN--A radio source.
\item[2]Coincidences, above the chance level, with known Active Galactic Nuclei (AGN) out to 75 Mpc.
\end{itemize}

Their conclusion that the primaries are protons is based on the contention that the deflections in the intergalactic medium (IGM) and the Galaxy would nullify the coincidences.

Although not stated, the need for a change in the Nuclear Physics follows from examination of the world's data (and their own – e.g. \cite{[6],[8]}) on the depth of shower maximum, which indicate
$\langle \ln A \rangle \sim 1.4$ at 10 EeV and $\sim 2.5$ at 40 EeV, the highest energy point plotted in the Auger results. With the conventional Nuclear Physics model, protons ($\langle \ln A \rangle = 0$ ) are certainly ruled out for the particles above 56 EeV. If true, this result would arguably be more important than the demonstration that AGN are responsible for the ultra high energy particles. 

This, then is the problem addressed here: 
Are 'heavy nuclei' ($\langle \ln A \rangle = 2 \div 3$) ruled out? 

\section{The Ankle} 
As remarked already, and referred to by us in several publications (eg. \cite{[7]}) we consider that this feature marks the transition from a mainly Galactic (G -) to a mainly Extragalactic (EG -) origin. Some others have it as a property of EG – protons and a demonstration that this is the case would clearly support the Auger contention. We have made many arguments against the EG – protons/ankle hypothesis (eg. \cite{[7]}) and these are strengthened from observation of the Auger energy spectrum reported in \cite{[8]}. The ankle is so sharp as to make its explanation in terms of E - p quite untenable.

\begin{figure}
\centerline{
\includegraphics[width=7.5cm]{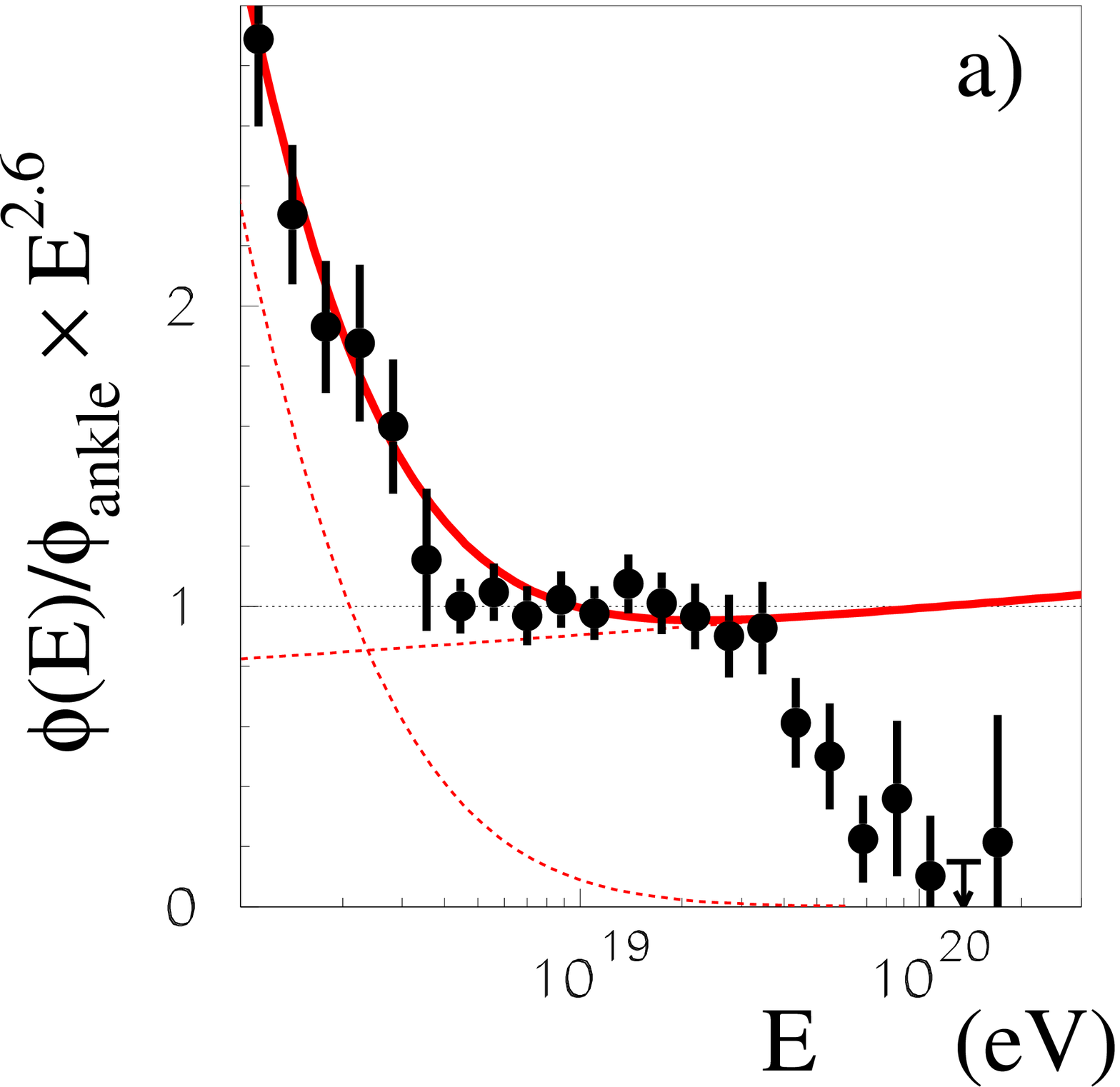} \hfill
\includegraphics[width=7.5cm]{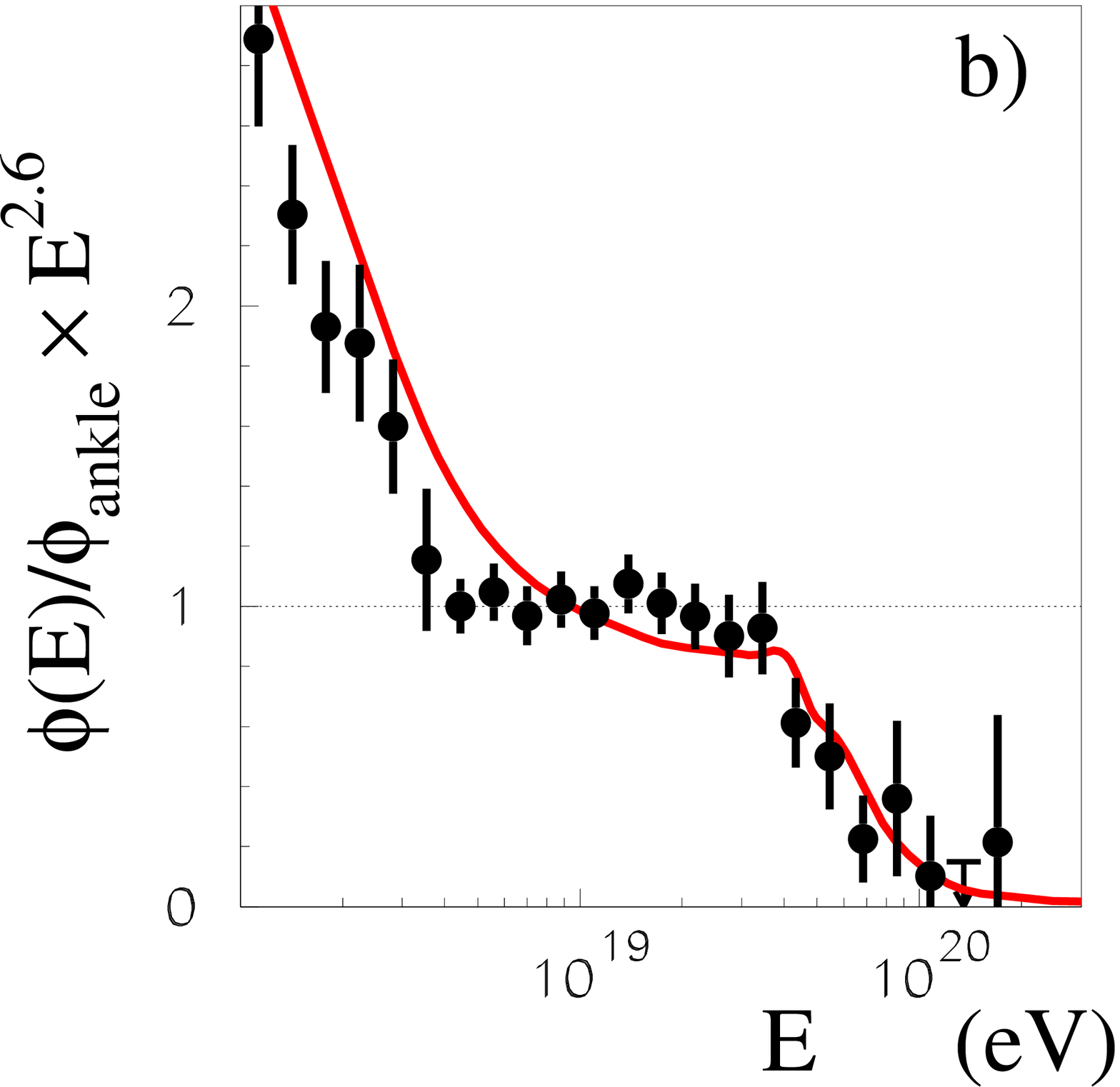}}
\includegraphics[width=7.5cm]{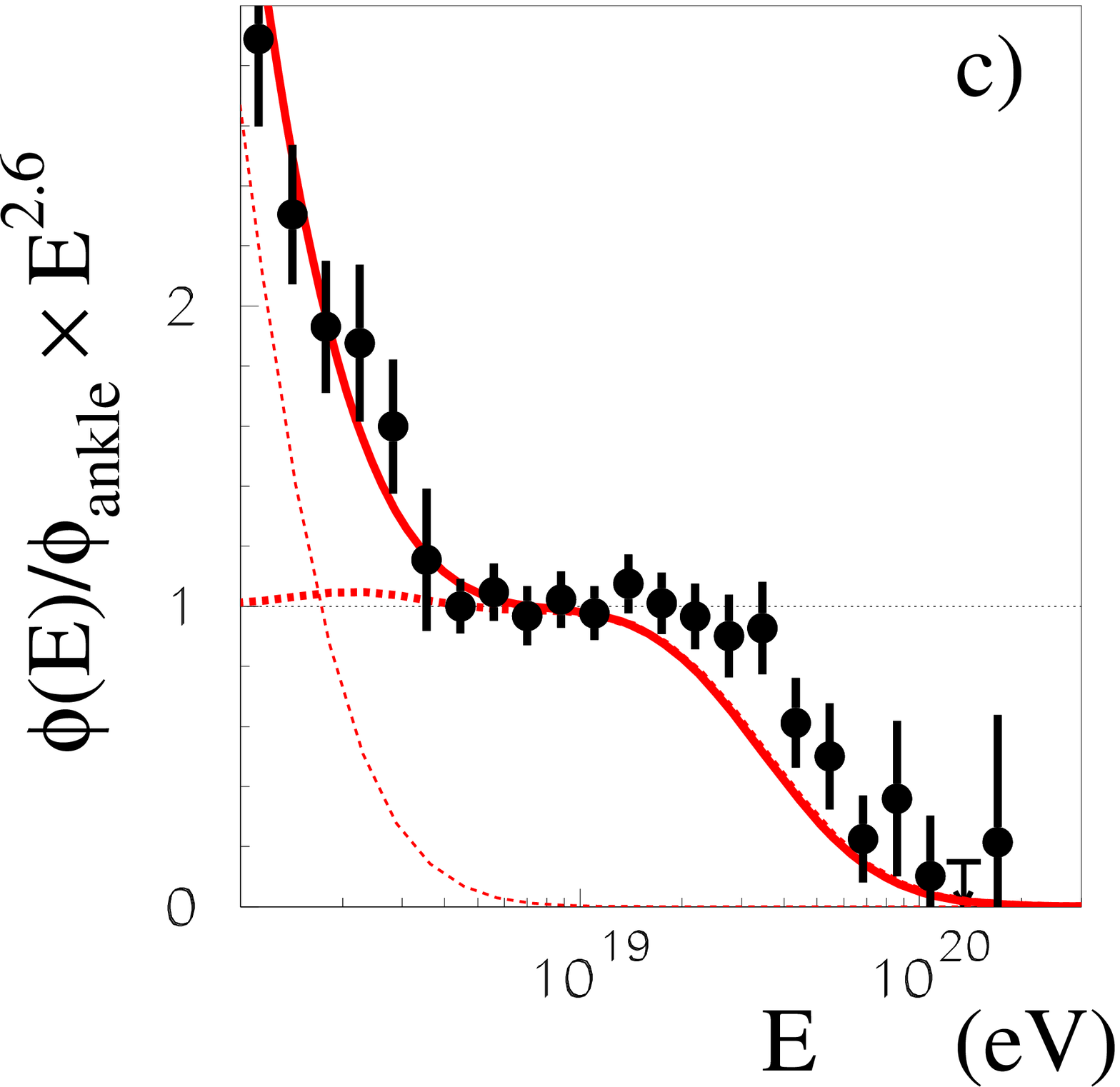}
\caption{The AUGER energy spectrum \cite{[8]} 
in comparison with various predictions\\
a. Our model fit \cite{[7]} 
where the Galactic (G -) and Extragalactic (EG -) spectra are simple power laws.\\
b. Comparison with EG – protons only model of \cite{[4]}.\\
c. Comparison with our Q2 model for protons \cite{[7]}. 
The shortage above 10 EeV would be covered by heavier nuclei. \label{fig2}}
\end{figure}

The situation can be seen by reference to Figure~\ref{fig2}. A two – component spectrum with the spectra Galactic (G -) and Extragalactic (EG -) crossing sharply, as in Figure~\ref{fig2}a, clearly gives a sharp ankle. The EG – protons alone, spectrum (Figure~\ref{fig2}b) has too smooth a transition; in fact, various factors would make the transition smoother still for such a model \cite{[7]}. Figure~\ref{fig2}c refers to one of our variants \cite{[7]}; specifically, origin of ultra – high energy cosmic rays (UHECR) in quasars. It is our contention that the actual form of the spectrum can be used to define the best fit spatial distribution of the sources as a function or red shift $Q(z)$ – see \cite{[7]}. The rather sharp fall in the Auger spectrum at about 40 Eev has relevance here, as will be discussed in more detail elsewhere. At present a possibility is that heavy nuclei are involved,indeed,this is the basis of our following arguments.In fact the prediction shown,which is for protons alone would need to be displaced downwards to allow for 'heavy' nuclei at  energies below 20 EeV,too.  

\section{The Depth of Shower Maximum}

\begin{figure}
\centerline{
\includegraphics[width=8.5cm]{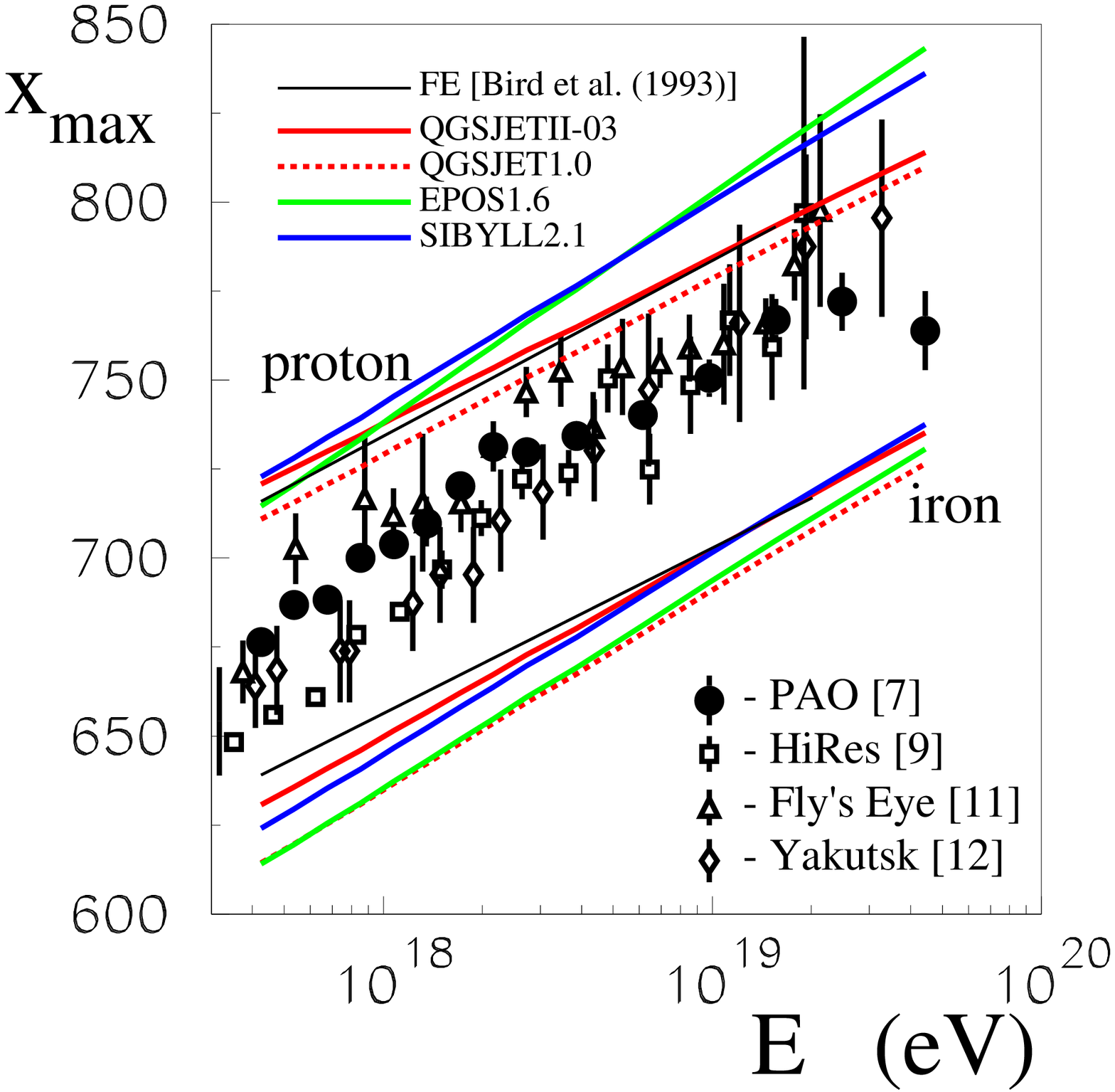}}
\caption{Depth of shower maximum versus energy measured by different experiments compared with different model predictions. 
\label{fig3}}

\end{figure}

\begin{figure}[bh]
\centerline{
\includegraphics[width=9cm]{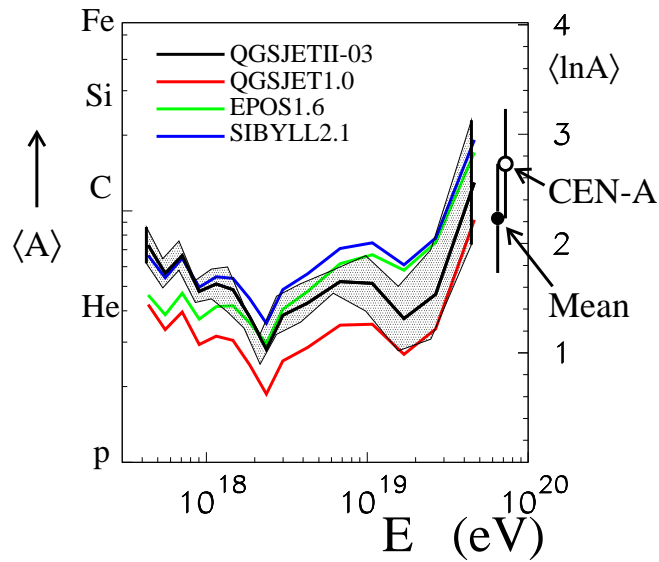}}
\caption{$\langle \ln A \rangle$ vs energy from our analysis of the AUGER results and different models. Most other $X_{\rm max}$ values from Figure~\ref{fig3} would give higher mean masses.
\label{fig4}}
\end{figure}

Measurements over many years have shown that the depth of maximum increases with increasing energy and its value is roughly mid – way between expectation for 'all protons' and 'all iron' for essentially all the nuclear models to date. With their superior statistics and analysis, the Auger work \cite{[8]} has shown that there is structure in the energy dependence, with a feature near the ankle
 energy. Figure~\ref{fig3} shows the results and Figure~\ref{fig4} shows the resulting $\langle \ln A \rangle$ from our analysis. It is interesting to note that the Hi–Res EAS array shows a similar feature: an $X{\rm max}$ change close to the ankle \cite{[9]}. A relevant matter to consider now is the expected mass composition {\it before} the ankle. This cannot be anywhere near mainly 'protonic' because of the lack of large anisotropies favouring the Galactic Plane, as shown by us in a previous detailed analysis \cite{[10]}. Thus, the Nuclear Physics models should not be too inaccurate here. Were the particles to be mainly protonic only above the ankle, the change in Nuclear Physics Model would need to take place over half a decade of energy, at most, and we consider this to be unphysical. 

At this stage, it can be remarked that, in fact, the Auger $X{\rm max}$ results would give too high an anisotropy at 1EeV,where the particles are of Galactic origin. The lower $X{\rm max}$ values measured by most others would not \cite{[10]}. 

\section{The arrival directions}
\subsection{CEN--A}
Returning to Figure~\ref{fig1}, together with the Auger authors, we are impressed by the signal from Centaurus-–A (CEN–-A), a long favoured source with its double jet, high power and flat radio spectrum. We would argue that the 8 particles within 20\% of CEN–-A are probably due to it. The lack of better symmetry about the source can be understood in terms of the varying collection efficiency in this region together with lack of symmetry in the magnetic fields in the IGM. We call this region A. Two more regions may be populated by specific nearby sources, denoted B and C in the Figure. 

Although it is claimed that there is an excess above chance of coincidences with AGN in general the statistics will be made worse when the CEN–-A events are removed, and, furthermore, there is a remarkable lack of events in the region above CEN-–A, where the number of AGN is seen to be very large. This approaches the VIRGO region (already mentioned), where the cluster is situated 17 Mpc away. It is instructive to make an estimate of how many particles might have been expected to be seen by Auger from VIRGO.
   Including the difference in collection efficiency, a factor $\sim 2.5$, we would expect to see, for a single CEN-–A source at the distance of VIRGO,  about 0.3 events,therefore there are less than a few CEN--A type   galaxies  amongst several thousand galaxies, and probably $\sim 10$ AGN in that cluster. Were all AGN like CEN–-A we would expect to see $\sim 3$ events here. This, not unexpected, variability of output in CR amongst AGN of different types coupled with the lower detection efficiency of distant AGN – and not to mention their time variability – tends to give further doubt to an analysis in which coincidences are sought between a non – homogeneous set of AGN and UHECR. In fact, there was a prior likelihood of large radio – galaxies rather than AGN in general – being likely sources or UHECR for two reasons:                                                                                                         
\begin{itemize}       
\item[a.]the likely detection of M87, a radio galaxy with a pronounced jet in the Northern hemisphere, \cite{[3]} and,
\item[b.]the obvious need for a type of source with large linear dimensions and, preferably radio spectra with small exponent indicating, for electrons at least, flat energy spectra. 
\end{itemize}
For radio galaxies, those giving the highest radio fluxes at earth are, in order of increasing distance from earth:
\begin{itemize}
\item
CEN-–A  (NGC 5128) at 4.9 Mpc
\item
VIRGO – A  (NGC 4486, M87) at 16.8 Mpc
\item
and FORNAX  (NGC 1316, ARP 154) at 16.9 Mpc
\end{itemize}

\subsection{'Source B'}
In the list just given, the first two have been mentioned already. FORNAX is seen to be not far from the 'Source B' but probably too far to be physically associated. However, it is in the FORNAX cluster and this has galaxies extending across to $l, b$: 200$^\circ$, $-40^\circ$. Most notable is that of the radio sources with flat radio spectra (associated with elliptical galaxies), \cite{[13]} the flattest, with exponents $\sim$0 and – 0.1, are near to Source B. They are NGC 1052 and NGC 1407 at $l$, $b$, distance: 182$^\circ$, $– 58^\circ$; 17.8 Mpc and 209$^\circ$, $- 50^\circ$, 21.6 Mpc.

We conclude that there are reasonable contenders for 'Source B'. 

\subsection{'Source C'}
The evidence for this 'cluster' of UHECR arrival directions being associated with a single known source is not strong. There are no obvious candidates. There are only a few 'normal' galaxies within 20 Mpc \cite{[14],[15]} although there is a nearby galaxy within 5 Mpc. Presumably a source further away is responsible?

\subsection{Other Source Complications}
Its well known that many AGN are time – variable (and CEN–-A is no exception). Thus, in view of transit time differences between UHECR and protons for very distant sources, the optical and UHECR sky may differ. It is interesting to note that nearby colliding galaxies (some of which go on to produce  AGN) have not (yet) been seen (see \cite{[15]} for previous work). In addition, to the different types of AGN their distances are clearly of great relevance; thus, catalogues are needed of putative claims for coincidences giving particle energies and AGN distances (and types).  
{\bf It can be remarked at this  stage that the term 'Active Galactive Nuclei' is probably a misnomer to describe the UHECR sources.The large radio sources have ceased to have active nuclei by the time the radio jets are seen.}

\section{The case for, or against, non protons}
\subsection{Acceleration Mechanisms}
Starting with acceleration, there is an obvious advantage in accelerating high – $Z$ particles insofar as the commonly considered acceleration mechanisms operate, with a rapidly falling energy spectrum, to some maximum rigidity. Thus, provided there is an adequate pool of pre-–accelerated nuclei, say up to iron, such non – protons should be at an advantage. There is a similar situation in the tens of thousands of GeV region, where the mean mass increases with energy \cite{[16]}.

\subsection{Angular deflections in the Galaxy}
In \cite{[1]} it is claimed that the deflections in the Galactic magnetic field do not allow the incoming nuclei to maintain their common direction. There are two points that should be made here: 
\begin{itemize}
\item[a.] For CEN–-A  the trajectories are very largely \underline{along} the coherent magnetic field direction (the Sagittarius Arm).
\item[b.] Calculations for the random field, using the parameters of field strength and reversal length \cite{[17],[18]}, yield a median deflection of 0.72 degrees, again for CEN–-A
\end{itemize}
Turning to 'Source B' at $l \sim 190^\circ$, $b \sim 60^\circ$, the direction gives paths mainly in our weak 'spur'.

Finally, for 'Source C' at $l \sim 60^\circ$, $b \sim 45^\circ$, again the direction is along a spiral arm, specifically along the inter arm region between the Orion and Sagittarius arms – where the magnetic field is lower, as well as being mainly ineffectual because of direction. Furthermore, the length of path through the irregular field is smaller than that for CEN–-A (by a factor sin $\sin 19^\circ/ \sin 45^\circ = 0.46$).

\subsection{Angular deflections in the IGM}

\begin{figure}
\centerline{
\includegraphics[width=7.5cm]{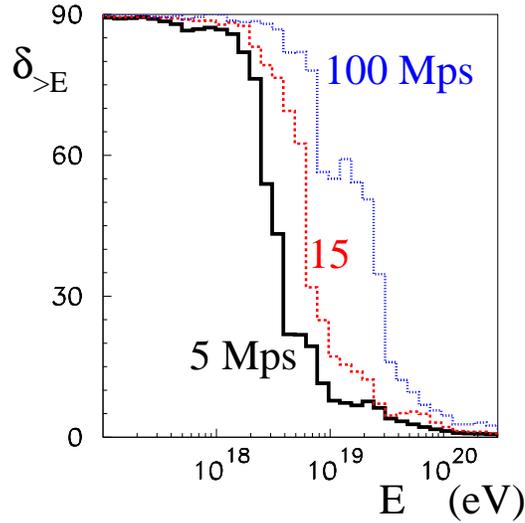}}
\caption{Integral distribution of angular deflections ($\delta$ - in degrees) for protons from 
CEN--A (line labeled 5~Mpc) and the distributions for 15~Mpc and 100~Mpc distant sources for a comparison.
\label{fig5}}
\end{figure}

The problem of calculation is more severe here, largely through uncertainty in the reversal length of the field, or to be more specific, the power spectrum of the irregularities \cite{[19]}. Elsewhere \cite{[15]} we gave an approximate expression for the root mean square angular deflection
$510 B \sqrt{(da)\;(pc)}$ degrees,
where $d$ is the distance to the source in Mpc, $a$ is the reversal length in Mpc and $pc$ is the energy in units of 1000 EeV. $B$ is in the field in $\mu$G. Here, we have $pc \sim$ 0.7. The result is similar to what we find using our more accurate method; the actual integral distribution is given in Figure~\ref{fig5} for the case of CEN-–A ($d$ = 5~Mpc); here, the median deflection is $1.1^\circ$ for protons. 

\section{Application to the UHECR Map}
Our method is to use the order of magnitude values of the median angular deviations from Figure 1, to give the expected median values of $Z$. Converting the mean value for the sources to an effective mass ($A = 2 Z$) gives our estimated $\langle \ln A \rangle$. 
This is then indicated on Figure~\ref{fig4}.

The value for CEN–-A, the best identified source, is seen to be $\langle Z \rangle = 7.7$ and 
$\langle \ln A \rangle$ follows as $\sim2.7$. Taking the mean of all three gives $\langle Z \rangle$ = 4.7 and $\langle \ln A \rangle = 2.2$. It seems to us unlikely that the true value is outside then limits; certainly, $\langle \ln A \rangle$ = 0 appears not to be needed. 

\section{Conclusions}

We conclude that it is probably not necessary to change the Nuclear Physics of high energy interactions at energy above 60 EeV, or so.

The way forward in the analysis of the Auger results is to endeavour to check the hypothesis that 'nearby' (within some 10s of Mpc) flat – spectrum radio galaxies are responsible. Identification will clearly rely on examination of the allotted energies to events within clusters as a function of radial distance from the possible source. Individual $X_{\rm max}$ – values need treating in the same manner.  

A complication, affecting all searches, is the fact that the
distant source may not be seen optically to be 'still on' when the particles arrive, \cite{[20]}.

\begin{table}
\caption{
Median expected displacements (in degrees) for protons from the sources indicated, and their 'total', i.e. addition in quadrature. Comparison with observed displacements gives an order of magnitude estimate of the particle charge, $Z$. \label{table}}

\vspace{.5cm}
\begin{tabular}{|l|c|cc|c|c|c|}
\hline
Source   &  Distance (Mpc) & Galaxy & IGM &Total & Median displacement observed&$Z$\\
\hline
CEN--A     &5 &0.7 &1.1 &1.3 &10 &7.7\\
Source B&20  &0.46 & 2.2&2.2&6&2.7\\
Source C&33&0.48&2.8&2.8&10&3.6\\
\hline
\end{tabular}
\end{table}


\section*{References}
\begin{thebibliography}{10}
\bibitem{[1]}{\it Auger Collaboration}, Science  {\bf 318}  938 (2007). 
\bibitem{[2]}Watson, A.A., QJL, R.Astro. Soc, {\bf 21}, 1 (1980).
\bibitem{[3]}Szabelski, J., Wdowczyk, J. and Wolfendale, A. W., J. Phys.G. {\bf 12}, 1433 (1986).
\bibitem{[4]}Berezinsky, V.S and Grigorieva, S.I, Astron. Astrophys. {\bf 199}, 1, (1998).
\bibitem{[5]}Chi, X., Wdowczyk, J. and Wolfendale, A.W., J.Phys.G {\bf 18}, 1869 (1992).
\bibitem{[6]}Wibig,T. and Wolfendale, A.W., J.Phys.G, {\bf 25}, 1099 (1999).
\bibitem{[8]}Watson, A.A, Proc 30th ICRC (Merida) (2007). 
\bibitem{[7]}Wibig, T. and Wolfendale, A.W, J. Phys, G. {\bf 34} 1891 (2007).
\bibitem{[9]}Abu-Zayyard, T. {\it et al.}, Phys. Rev. lett., {\bf 84}, 4276 (2000).
\bibitem{[10]}Wibig, T. and Wolfendale, A. W., J. Phys, G. {\bf 25}, 2001 (1999).
\bibitem{[11]}Bird, D. J. et al., Ap. J., {\bf 424} 491 (1994).
\bibitem{[12]}Knurenko, S. et al., Proc. 17th ICRC (Hamburg) {\bf 1}, 177 (2001).
\bibitem{[13]}Hummel, E., {\it The Radio Continuum Structure of Bright Galaxies at 1.4~GHz}, PhD Thesis, Groningen (1980).
\bibitem{[14]}Tully, R. B., {\it Nearby Galaxies Catalog}, C U P (1988).
\bibitem{[15]}Al–-Dargazelli S. S. {\it et al.}, J Phys G, {\bf 22}, 1825 (1996).
\bibitem{[16]}Ogio, S., Proc.28th ICRC, {\bf 1}, 131 (2003).
\bibitem{[17]}French, D. K. and Osborne, J. L, MNRAS. {\bf 177} 569 (1976).
\bibitem{[18]}Giller, M. and Wolfendale, A. W., J.Phys.G. {\bf 19}, 449 (1993).
\bibitem{[19]}Wibig, T., Central European J. Phys. {\bf 2}, 277 (2004).
\bibitem{[20]}Wibig, T. and Wolfendale, A.W., J.Phys.G,
{\bf 30}, 524 (2004).
\end{thebibliography}
\end{document}